# A Causal Inference Approach to Measure the Vulnerability of Urban Metro Systems

Nan Zhang
nan.zhang16@imperial.ac.uk

Daniel J. Graham[*]
d.j.graham@imperial.ac.uk

Daniel Hörcher
d.horcher@imperial.ac.uk

Prateek Bansal
prateek.bansal@imperial.ac.uk

Transport Strategy Centre, Department of Civil and Environmental Engineering
Imperial College London, London, UK

[*]Corresponding author

## Abstract

Transit operators need vulnerability measures to understand the level of service degradation under disruptions. This paper contributes to the literature with a novel causal inference approach for estimating station-level vulnerability in metro systems. The empirical analysis is based on large-scale data on historical incidents and population-level passenger demand. This analysis thus obviates the need for assumptions made by previous studies on human behaviour and disruption scenarios. We develop four empirical vulnerability metrics based on the causal impact of disruptions on travel demand, average travel speed and passenger flow distribution. Specifically, the proposed metrics based on the *irregularity in passenger flow distribution* extends the scope of vulnerability measurement to the entire trip distribution, instead of just analysing the disruption impact on the entry or exit demand (that is, moments of the trip distribution). The unbiased estimates of disruption impact are obtained by adopting a propensity score matching method, which adjusts for the confounding biases caused by non-random occurrence of disruptions. An application of the proposed framework to the London Underground indicates that the vulnerability of a metro station depends on the location, topology, and other characteristics. We find that, in 2013, central London stations are more vulnerable in terms of travel demand loss. However, the loss of average travel speed and irregularity in relative passenger flows reveal that passengers from outer London stations suffer from longer individual delays due to lack of alternative routes.

Key words: vulnerability, urban metro system, causal inference, propensity score matching

# 1. Introduction

Metros, also known as subways or rapid transit, have become a vital component of public transport. With the advantage of large capacity and high-frequency services, 178 metro systems worldwide carried a total of 53,768 million trips in 2017 (International Union of Public Transport, 2018). Incidents occur frequently in urban metro systems, mainly due to supply-side failures (e.g., signal failures), sudden increase in travel demand (e.g., public concert or football matches) and change in weather conditions (Brazil et al., 2017; Melo et al., 2011; Wan et al., 2015). These incidents can cause service delays and overcrowding, which in turn lead to safety concerns and potential losses in social welfare. For instance, the London Underground encountered 7973 service disrupting incidents of above 2 minutes duration between April 2016 and April 2017, causing a total loss of around 34 million customer hours (Transport for London, 2017; Transport for London, 2019). The Singapore Mass Rapid Transit experienced 47 severe delays that lasted over 30 minutes between 2015 and 2017 (Land Transport Authority, 2017).

Operators may consider investing in new technologies to improve metro facilities and mitigate the effect of incidents. For instance, the New York City Subway was in a state of emergency in June 2017 after a series of derailments, track fires and overcrowding incidents. The Metropolitan Transportation Authority invested over $8 billion to stabilise and modernise the incident-plagued metro system (Metropolitan Transportation Authority, 2019). It is apparent that metros are willing to invest in their infrastructure systems, but it is often not known how those investments compare in achieving improvements. To facilitate project selection, metros are increasingly relying on disaggregate performance metrics that reveal the most vulnerable parts of the network. Performance can be measured in various ways. Popular examples are risk, resilience, reliability and vulnerability related metrics. These concepts are often confused by researchers as well as well as practitioners. Interested readers can refer to Faturechi and Miller-Hooks (2015) and Reggiani, Nijkamp and Lanzi (2015) to understand the most agreed relationship among these concepts. In this paper, we focus on the vulnerability of urban metro systems, where the performance measures of interest are passenger demand, average travel speed and passenger flow distribution.

Since the 1990s, the concept of vulnerability has been widely used to characterise the performance of transport systems (Mattsson and Jenelius, 2015; Reggiani, Nijkamp and Lanzi, 2015), which is often defined as a measure of susceptibility of the transport system to incidents (Berdica, 2002; Jenelius et al., 2006; O'Kelly, 2015). In this study, the vulnerability of metro systems refers to the extent of degradation in the level of service due to service disruptions. Service disruptions are defined as events that interrupt normal train operations for a specific period of time[1]. Disruptions should be distinguished from the broader term "incidents", as incidents might not always affect services. Examples of such incidents include elevator failure or corridor congestion in metro stations. Vulnerability metrics can measure the consequences of service interruptions, in the form of performance outputs such as train kilometres, passenger volumes or the quality of travelling. For operators, such metrics have important implications in identifying weak stations or links in metro systems and efficiently allocating resources to the most affected areas. Given the rising interest in utilising vulnerability metrics in disruption prevention and management, obtaining an accurate measure of such metrics is crucial.

---

[1] Five minutes to ten minutes are commonly used thresholds to define disruptions. Different metro systems around the world adopt several thresholds, primarily on the basis of the regular frequency of operations.

Traditionally, vulnerability in urban metros is investigated based on complex network theory and graph theory. Complex network theory converts metro networks into graphs, which enables the quantitative measurement of vulnerability in metro systems (Chopra et al, 2016; Derrible and Kennedy, 2010; Yang et al., 2015). The adoption of graph theory has facilitated the evolution of vulnerability indicators from simply capturing the characteristics of network topology to also considering travel demand patterns and their land use dependencies (Jiang et al., 2018). However, most of these studies rely on simulation-based approaches to quantify vulnerability under hypothetical scenarios of disruptions. These simulation experiments are based on assumptions, both in terms of passenger behaviour and the type and scale of disruptions (Lu, 2018; Sun and Guan, 2016; Sun et al., 2015; Sun et al., 2018). With an empirical approach, such assumptions can be avoided, and thus more reliable metrics of vulnerability can be achieved using historical evidence.

The empirical approach is rare but not unique in the literature. The exception we are aware of is Sun et al. (2016), who first detect incidents based on abnormal ridership and use the *real* incidents data to assess the vulnerability of the metro system. However, their method has some limitations. First, they assume the occurrence of incidents to be random, which is a strict and unrealistic assumption as we demonstrate in this study. Also, the abnormal ridership may not be a good indicator of incidents if the fluctuation in ridership are merely manifestations of changes in travel demand due to external factors.

This paper proposes a novel alternative methodology to quantify vulnerability, by *empirically estimating the causal impact* of service disruptions on travel demand, average travel speed and passenger flow distribution at station-level. The application of a propensity score matching method accounts for the non-randomness of disruptions and ensures unbiasedness of the causal estimates. We make this approach comprehensive for the entire network, including stations where disruptions are not observed, by predicting the level of vulnerability at these stations with a random forest algorithm. In this way, we eliminate the need for ad hoc assumptions on passenger behaviour and the nature of disruptions.

We use London Underground as a case study and apply the methodology with large-scale automated fare collection and incident data. The station-level vulnerability is heterogeneous among the network, depending on the considered performance metrics. In terms of the demand loss and gross speed loss (overall delay), the most affected stations are more likely to be found in Central London areas. When considering average speed loss (individual delay) and irregularity in relative passenger flows, the most affected stations are scattered around outer London areas due to lack of alternative routes. These results can potentially aid investment decisions of metro operators.

The rest of paper is organised as follows. Section 2 reviews the literature on vulnerability measurement and disruption impact analysis in urban metro systems. Section 3 presents our empirical framework to compute vulnerability metrics. This section discusses the proposed causal inference approach to estimate the unbiased disruption impact, which is the key input in building vulnerability metrics. In Section 4, we analyse the vulnerability of London Underground as a case study. Results are discussed in Section 5. Finally, Section 6 concludes and highlights the potential avenues for future research.

## 2. Literature review

Below we provide a contextual review of previous studies related to vulnerability measurement. In Section 2.1, we review the literature on vulnerability quantification in rail transit networks, while Section 2.2 investigates previous attempts to estimate the impact of disruptions.

**2.1 Measuring the vulnerability of metro systems**

There are two traditional methods used to build vulnerability indicators of metro systems – topology-based and system-performance-based analysis.

The topological methods rely on complex network theory to convert the metro network into a scale-free graph, in which nodes represent metro stations, edges represent links between directly connected stations and the weight associated with each edge is computed based on travel time or distance (Derrible and Kennedy, 2010; Mattsson and Jenelius, 2015; Zhang et al., 2011). The changes in the system's connectivity are reflected on graphs by removing nodes or links and vulnerability is entirely governed by the topological structure. For instance, the location importance of metro stations or links is indicated by the number of edges connected to a specific node and the fraction of shortest paths passing through the given node/edge (Sun and Guan, 2016; Sun et al., 2018; Yang et al., 2015; Zhang et al., 2018). Network-level efficiency is indicated by the average of reciprocal shortest path length between any origin-destination (OD) pair. Such global indicators capture the overall reachability as well as the service size of a metro system (Sun et al., 2015; Yang et al., 2015).

System-performance-based analyses not only consider the network topology but also incorporate real data on metro operations (e.g., ridership distribution) into vulnerability measurement (M'cleod et al., 2017; Mattsson and Jenelius, 2015). For instance, Sun et al. (2018) use a ridership-based indicator – a sum of flows in edges connected with the given node – to complement the topological measures by integrating passengers' travel preferences. Other studies use passenger delay and demand loss as vulnerability indicators (Adjetey-Bahun et al., 2016; M'cleod et al., 2017; Nian et al., 2019; Rodríguez-Núñez and García-Palomares, 2014). Specifically, passenger delay is summarised by changes in the weighted average of travel time between all OD pairs due to disruptions where weights are station-level passenger loads. Jiang et al. (2018) suggest integrating land use characteristics around stations into vulnerability measurement because metro systems interact with the external environment during incidents.

To quantify vulnerability based on the aforementioned indicators of the system's performance, almost all previous studies adopt simulation-based approaches and assume hypothetical disruption scenarios. The simplest disruption scenario involves a single station or link closure, assuming one node or edge in the graph is out of service. This incident affects the topology structure and passengers' route choice and the differences in the corresponding performance indicators under normal and disrupted scenarios are quantified to measure vulnerability (Sun et al., 2015). More complex disruption scenarios include the closure of two or more non-adjacent stations, failure of an entire line, and sequential closure of stations until the network crashes (Adjetey-Bahun et al., 2016; Chopra et al., 2016; Sun and Guan, 2016; Zhang et al., 2018; Zhang et al., 2018). Ye and Kim (2019) also discuss the case of partial station closure.

Simulation-based studies gained popularity because they do not require incident data and can flexibly control simulation settings to imitate a wider range of possible situations. However, researchers have to make many assumptions to infer passengers' response to virtual disruptions. Without observing passengers' movements during real incidents, the validity of the simulation assumptions is questionable. For example, while quantifying passenger delay indicators, Rodríguez-Núñez and García-Palomares (2014) and Adjetey-Bahun et al. (2016) assume that all passengers have the same travel speed and they do not change their destinations under disruptions unless there is no available route. However, in reality, passengers can travel at different speeds, leave the metro system, change their destinations, or reroute during disruptions. As a result, especially for system-based analyses, vulnerability metrics obtained from simulation-based studies may not reflect the true changes in the level of service due to disruptions. There is, therefore, scope to improve vulnerability measurement by empirically estimating the impact of disruptions. The advantage of empirical-based methods is that the aforementioned assumptions are no longer needed, and the estimated impacts of disruptions are more reliable. However, the need for large-scale datasets is the main drawback of empirical studies.

**2.2 Estimating disruption impact**

In an urban rail transit context, early attempts to analyse disruption impact relied on surveys. Rubin et al. (2005) conducted a stated preference survey to understand the psychological and behavioural reactions of travellers to the bombing incident, which happened in London during July 2005. They consider passenger's reduced intention of travelling by the London Underground after the attack as the key indicator. Since stated willingness may not reflect real travel behaviour, Zhu et al. (2017) performed a revealed preference survey to investigate travellers' reactions to transit service disruptions in Washington D.C. Metro. By comparing their actual travel choices before and during the metro shutdown, they find a 20% reduction in demand. Results from such surveys are usually presented as the percentage change in passengers' preferences for travel modes, departure time, and destinations. Although this information is useful, we still need detailed information about delays or demand losses to quantify true disruption impacts. Furthermore, there are inherent limitations of survey-based studies. For instance, repeated observations of a respondent are difficult to collect for a long period because of constraints associated with cost, manpower, recording accuracy, and privacy protection of respondents (Kusakabe and Asakura, 2014). A survey sample also cannot cover all passengers, which may lead to biased estimates of disruption impact if the sample is not representative of the population.

With the wide use of automated fare collection facilities in metro systems, smart card data have become a powerful tool for research related to transit operations and travel behaviour (Pelletier, Trépanier and Morency, 2011). Compared to survey data, the key advantages of smart card data are cost-effectiveness, continuous long-term recording and accurate travel information for each passenger within the system (Kusakabe and Asakura, 2014). Therefore, researchers have started using smart card data to analyse disruption impacts. For instance, Sun et al. (2016) develop a method to identify incidents and conduct trip assignments with/without incidents. They estimate the disruption impact by computing the differences between two assignments in terms of ridership distribution and travel time across all OD pairs. This study does not require extra assumption about passengers' reaction because their actual locations and movements are revealed from smart card data. However, they assume that metro disruptions occur randomly, while in reality, factors such as travel demand, signalling type, passenger behaviour, operating years, rolling stock characteristics and weather conditions have a significant influence on the likelihood of metro failures (Brazil et al., 2017; Melo et al., 2011; Wan et al., 2015). This is a particularly important consideration because the impact estimated from direct comparison of performance indicators before and after disruptions will be biased under non-random occurrence of disruptions. Specifically, a few factors affecting the impact of disruptions (e.g., passenger behaviour and weather conditions) may also affect the occurrence of disruptions, leading to confounding bias in pre-post comparison estimates (Imbens and Rubin, 2015). Some researchers also adopt prediction-based approaches to quantify disruption impact using smart card data. For instance, Silva et al. (2015) propose a framework to predict the exit ridership and model behaviours of passengers under station closure and line segment closure. In a very recent study, Yap and Cats (2020) apply supervised learning approaches to predict the passenger delay caused by incidents. However, these prediction-based studies also cannot disentangle the causal effect of disruptions and can result into biased estimates due to the existence of confounding factors.

Table 1 shows a comparison of recent vulnerability studies and also illustrates the contribution of this research. We conclude this section with a summary of gaps in the literature that we address to obtain more accurate measures of vulnerability:

1. Previous studies on vulnerability metrics of transit systems are largely based on simulation approaches. These studies do not account for the actual behaviour of passengers under disruptions. Basing analyses on empirical data, rather than simulations, obviates the need for making potentially unrealistic assumptions on passengers' movement.

2. In urban metro systems, disruption occurrences can be non-random. Therefore, empirical studies on quantifying disruption impacts should account for this non-randomness to eliminate confounding biases in estimation.

In this paper, we show that both improvements can be made by adopting causal inference methods and calibrating them using large-scale smart card data and incident data. Specifically, the proposed method allows for the non-random occurrence of disruptions and adjusts for potential bias caused by confounding factors. Subsequently, unbiased empirical estimates of disruption impact are used to accurately compute vulnerability metrics of metro systems.

**Table 1:** A comparison of recent research on metro vulnerability.

| Research | Vulnerability metrics or disruption impacts | | Analysis approach | | Smart card or OD data | Land-use | Non-random disruptions |
|---|---|---|---|---|---|---|---|
| | Topology-based | System performance-based | Simulation-based | Empirical (real incidents) | | | |
| Derrible and Kennedy, 2010 | √ | | √ | | | | |
| Zhang et al., 2011 | √ | | √ | | | | |
| Yang et al., 2015 | √ | | √ | | | | |
| Chopra et al, 2016 | √ | | √ | | | | |
| Zhang et al., 2018 | √ | | √ | | | | |
| Zhang et al., 2018 | √ | | √ | | | | |
| Ye and Kim, 2019 | √ | | √ | | | | |
| Rodríguez-Núñez and García-Palomares, 2014 | | √ | √ | | √ | | |
| Adjetey-Bahun et al., 2016 | | √ | √ | | √ | | |
| M'cleod et al., 2017 | | √ | √ | | √ | | |
| Sun et al., 2015 | √ | √ | √ | | √ | | |
| Sun and Guan, 2016 | √ | √ | √ | | √ | | |
| Sun et al., 2018 | √ | √ | √ | | √ | | |
| Lu, 2018 | √ | √ | √ | | √ | | |
| Jiang et al., 2018 | | √ | √ | | √ | √ | |
| Sun et al., 2016 | | √ | | √ | √ | | |
| **Our approach** | | √ | | √ | √ | √ | √ |

## 3. Methodology

From a methodological point of view, our empirical approach has three stages: first, we apply a causal inference method to estimate the impact of disruptions on station-level travel demand and travel speed (see Section 3.1). Then, in Section 3.2. we construct vulnerability metrics based on the disruption impact estimated in the first stage. Finally, the third stage imputes[2] missing vulnerability metrics for non-disrupted stations using machine learning algorithms. Figure 1 illustrates all steps of the proposed empirical framework.

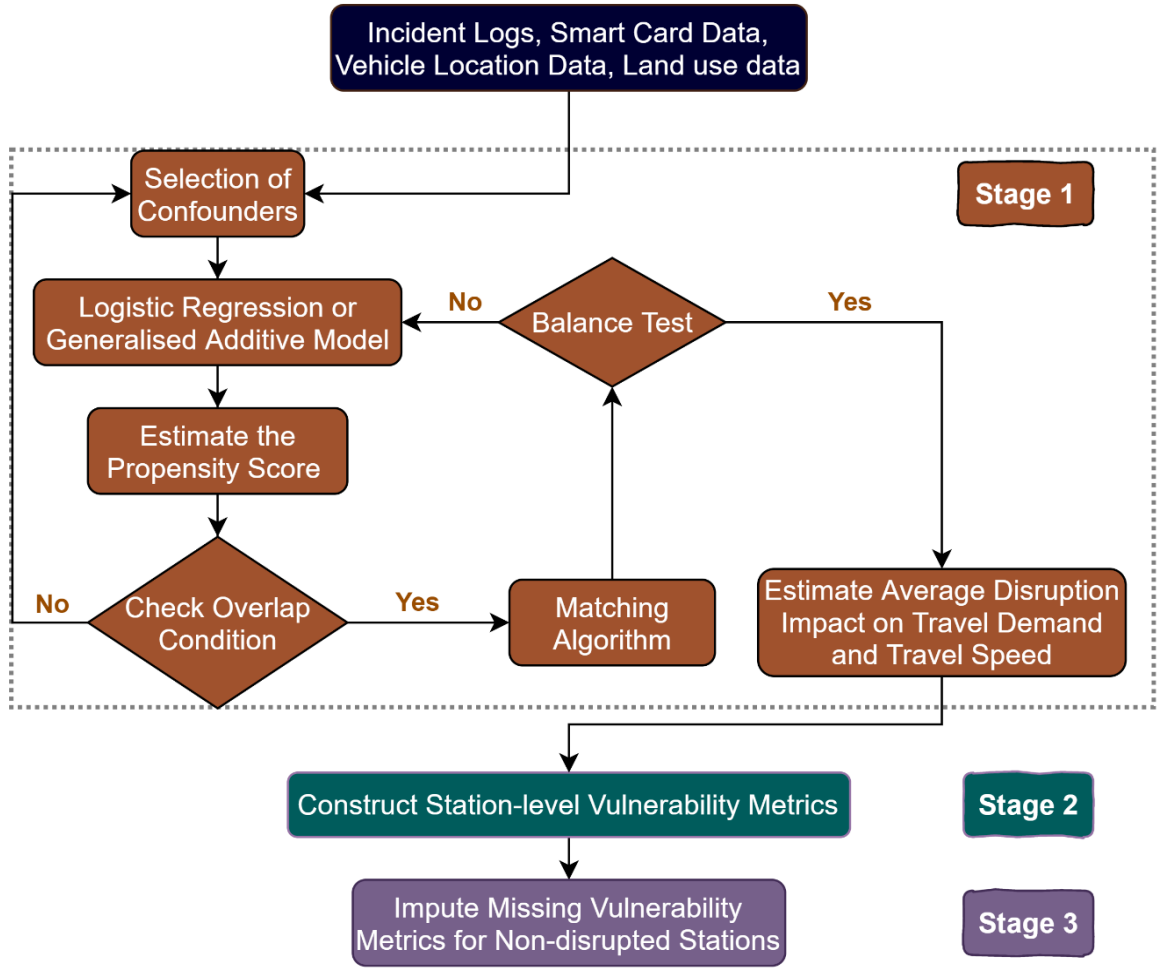

**Figure 1:** Flowchart of the paper's methodological framework.

### 3.1 Stage 1: Causal inference method to estimate disruption impact

To evaluate the impact of a disruption on a metro system, we use Rubin's potential outcome framework to establish causality (Rubin, 1974). We define metro disruptions as 'treatments' and the objective of our analysis is to quantify the causal effect of treatments on 'outcomes' related to system performance[3]. Specifically, we are interested in estimating station-level causal effects of disruptions on i) travel demand, ii) travel speed of passengers, and iii) passenger flow distributions from/to a station. From the literature, we know that factors such as passenger demand, weather conditions, network topology and engineering design influence the likelihood of disruption occurrence (Brazil et al., 2017; Melo et al., 2011; Wan et al., 2015). Therefore, the assignment of the treatment is not random. This is important in our context because the factors associated with the assignment of the treatment are also likely to affect the outcomes of interest, and are thus potential confounders in estimation of impacts. Since previous studies on disruption impact have ignored the non-randomness of treatments, their estimated impact may be biased.

---

[2] In Statistics, "imputation" is the process of replacing missing data with substituted values. Here we retrieve these missing values based on a relationship between vulnerability metrics and covariates of the disrupted stations.

We adopt propensity score matching (PSM) methods to address this issue, which potentially eliminates such confounding biases. The propensity score is defined as the conditional probability that a unit receives treatment given its baseline confounding characteristics. If the observed characteristics sufficiently capture the sources of confounding, then the propensity score can be used to consistently estimate impacts given conditional independence between treatment assignment and outcomes (e.g. conditional on the propensity score) (Imbens and Rubin, 2015). This index is obtained by estimating a relationship between treatment assignment and baseline confounding characteristics using a regression model. The estimated propensity score is then used to form various semi-parametric estimators of the treatment effect such as weighting, regression, and matching. In this section, we first provide a contextual formulation of PSM and then describe how we apply PSM to quantify the causal impact of metro disruptions on the performance of metro systems.

*3.1.1 Propensity Score Matching (PSM) Methods*

The system-level impact, which averages the impact of all disruptions occurred within the metro system, is too generic to represent network vulnerability. Thus, we focus instead on estimating station-level disruption impacts. We define study unit $i$ as the observation of a metro station within a 15-minute interval. The treatment variable, denoted by $W_{it} \in \{0, 1\}$, records whether study unit $i$ at time $t$ is observed in a disrupted ($W_{it} = 1$) or undisrupted state ($W_{it} = 0$). To quantify disruption impacts, we define outcomes of interest as the changed travel demand and average speed of trips that start from the given study unit, denoted by $Y_{it}$.

$$Y_{it}(W_{it}) = Y_{it}(0) \times (1 - W_{it}) + Y_{it}(1) \times W_{it} \qquad (1)$$

$$Y_{it} = \begin{cases} Y_{it}(0) & if\ W_{it} = 0 \\ Y_{it}(1) & if\ W_{it} = 1 \end{cases}$$

$$i = 1, \dots, n \quad t = 1, \dots, T,$$

where $n$ is the total number of stations within the metro system, and $T$ is the total number of time intervals during the study period (for example, T=4 if study period is 1 hour). $Y_{it}(0)$ and $Y_{it}(1)$ are counterfactual potential outcomes, only one of which is observed. The propensity score, denoted by $e(X_{it})$, is obtained by regressing $W_{it}$ on confounding factors, denoted by $X_{it}$. We discuss potential confounding factors in the empirical study in Section 4.

---

[3] In causal inference, 'treatment' means the intervention or exposure assigned to (or encountered by) study units, and 'outcomes' means the observed results or effects of the intervention on a response variable of interest. In the context of this study, service disruptions that occurred at metro stations are the 'treatment', and 'outcomes' are the performance of metro services such as travel demand, journey speed, and passenger flow distribution.

To derive valid causal inference using PSM we need our model to satisfy three key assumptions. The first one is the conditional independence assumption (CIA),

$$W_{it} \perp (Y_{it}(0), Y_{it}(1)) \mid X_{it}, \tag{2}$$

which states that conditional on the observed confounding factors $X_{it}$, the treatment assignment should be independent of the potential outcomes. The advantages of the propensity score stems from a property that this conditional independence can be achieved by just conditioning on a scalar rather than high-dimensional baseline covariates (Rosenbaum and Rubin, 1983). Thus, the CIA based on the propensity score can be written as:

$$W_{it} \perp (Y_{it}(0), Y_{it}(1)) \mid e(X_{it}). \tag{3}$$

The second assumption requires common support in the covariate distributions by treatment status:

$$0 < pr(W_{it} = 1 | X_{it} = x) < 1 \quad \text{for all } x, \tag{4}$$

which states that the conditional distribution of $X_{it}$ given $W_{it} = 1$ should overlap with that of the conditional distribution of $X_{it}$ given $W_{it} = 0$. This assumption can be tested by comparing the distributions of propensity scores between treatment and control groups.

The third assumption, also known as the stable unit treatment value assumption (SUTVA), requires that the outcome for each unit should be independent of the treatment status of other units (Graham et al., 2014).

If all three assumptions hold and the outcome variable is entry demand or travel speed, the average treatment effect (ATE) of disruptions on a station $i$ can be derived using the following equations (Imbens and Wooldridge, 2009):

$$\tau^i_{ATE} = \hat{\tau}^i_{match} = \frac{1}{T_d} \sum_{t=1}^{T_d} \left( \hat{Y}^i_t(1) - \hat{Y}^i_t(0) \right), \tag{5}$$

$$\hat{Y}^i_t(1) = Y_{it},$$

$$\hat{Y}^i_t(0) = \frac{1}{M} \sum_{t_c \in J_M(it)} Y_{it_c},$$

$$i = 1, \dots, n \quad t = 1, \dots, T_d,$$

where $t \in \{1, \dots, T_d\}$ denotes all the disrupted time intervals of station $i$ during the study period and $Y_{it_c}$ is the outcome of the control unit $t_c$ corresponding to station $i$ disrupted or treated at time $t$. $J_M(it)$ is a set of indices of the closest $M$ control units (in terms of propensity scores) for station $i$ disrupted at time $t$ during the same 15-minute interval, but on a different day[4]. Thus, $\hat{\tau}^i_{match}$ represents the average of the difference between the outcomes of treated and matched control units.

When the outcome variable is trip distribution, ATE can be expressed as:

---

[4] Please note that the study period of this study is 35 days. Therefore, we observe the same station across multiple days (see Section 4 for details).

$$\tau^i_{ATE} = \hat{\tau}^i_{match} = \frac{1}{T_d}\sum_{t=1}^{T_d}\left[dif\left(\hat{Y}^i_t(1),\ \hat{Y}^i_t(0)\right)\right], \tag{6}$$

$$\hat{Y}^i_t(1) = Y_{it} = \left(r^1_{1it},\ r^2_{1it},\ \ldots,\ r^k_{1it}\right),$$

$$\hat{Y}^i_t(0) = \frac{1}{M}\sum_{t_c \in J_M(it)} Y_{it_c} = \left[\frac{1}{M}\sum_{t_c \in J_M(it)}\left(r^1_{0it_c}\right),\ \ldots,\ \frac{1}{M}\sum_{t_c \in J_M(it)}\left(r^k_{0it_c}\right)\right],$$

$$i = 1, \ldots, n \qquad k = 1, \ldots, n \qquad t = 1, \ldots, T_d,$$

where for a treated or disrupted unit, $Y_{it}$ denotes the distribution of trips made from (outward) and to (inward) station $i$ at time $t$, $r^k_{1it}$ denotes the ridership from the disrupted station $i$ to station $k$ in case of outward flow (or from station $k$ to station $i$ in case of inward flow) at time $t$. Correspondingly, $Y_{it_c}$ denotes a composite distribution which averages the ridership distribution of all closest $M$ control units during the same 15-minute duration, but on a different day. $r^k_{0it_c}$ denotes the ridership between station $i$ and station $k$ for a non-disrupted period $t_c$ in the control group. $dif(a,b)$ is a function to calculate the distance between discrete distributions $a$ and $b$. In the context of this study, we consider three distance functions:

$$dif_1\left(\hat{Y}^i_t(1),\ \hat{Y}^i_t(0)\right) = \sqrt{\sum_{k=1}^n \left(r^k_{1it} - \frac{1}{M}\sum_{t_c \in J_M(it)}\left(r^k_{0it_c}\right)\right)^2}, \tag{7}$$

$$dif_2\left(P(\hat{Y}^i_t(1)),\ P\left(\hat{Y}^i_t(0)\right)\right) = \frac{1}{\sqrt{2}} \times \sqrt{\sum_{k=1}^n \left(\sqrt{P^k_{it}(1)} - \sqrt{P^k_{it}(0)}\right)^2}, \tag{8}$$

$$dif_3\left(P(\hat{Y}^i_t(1))\ \|\ P\left(\hat{Y}^i_t(0)\right)\right) = \sum_{k=1}^n \left[P^k_{it}(1) \times \log\left(\frac{P^k_{it}(1)}{P^k_{it}(0)}\right)\right], \tag{9}$$

$$P\left(\hat{Y}^i_t(1)\right) = \left(p^1_{it}(1), \ldots, p^k_{it}(1)\right)$$

$$P\left(\hat{Y}^i_t(0)\right) = \left(p^1_{it}(0), \ldots, p^k_{it}(0)\right)$$

$$p^k_{it}(1) = \frac{r^k_{1it}}{\sum_{k=1}^n \left(r^k_{1it}\right)},$$

$$p^k_{it}(0) = \frac{\frac{1}{M}\sum_{t_c \in J_M(it)}\left(r^k_{0it_c}\right)}{\sum_{k=1}^n \left(\frac{1}{M}\sum_{t_c \in J_M(it)}\left(r^k_{0it_c}\right)\right)},$$

where $dif_1(.)$ represents the Euclidean distance, which directly aggregates the difference between each element of the input distributions without normalising. The latter two functions compare the probability mass functions $P(\hat{Y}^i_t(1))$ and $P(\hat{Y}^i_t(0))$. $dif_2(.)$ represents the Hellinger distance and $dif_3(.)$ represents Kullback–Leibler divergence (also known as relative entropy). Each distance function has its strength and weakness, which we highlight in Section 5.4 while discussing results of the empirical study.

In the next subsection, we explain how the causal inference framework introduced in Equations (1), (5) and (6) can be implemented in the present application. Following the framework summarised in Figure 1, we first provide details of the propensity score model, followed by description of our matching algorithms and the estimation of disruption impacts.

*3.1.2 Application of PSM Methods*

To predict the propensity score, i.e. probability of encountering disruptions at a metro station within 15-minute interval conditional on the baseline confounding characteristics, we use the logistic regression model with a linear link function:

$$e(X_{it}) = pr(W_{it} = 1 | X_{it} = x^{\{c\}}) = p(it) \tag{10}$$

$$log\left[\frac{p(it)}{1 - p(it)}\right] = \alpha + \beta x^{\{c\}} \quad i = 1, \dots, n \quad t = 1, \dots, T,$$

where $\alpha$ is the intercept and $\beta$ is the vector of regression coefficients related to the vector of confounding factors $x^{\{c\}}$. In our empirical study, a station with a higher number of incidents in the past is more likely to encounter a new disruption in the future, just like the black spot on highways. To account for this temporal correlation among disruption occurrence, we ensure that confounding factors contain the history of past disruptions happened on the same day.

Additionally, we also consider a more advanced generalised additive model (GAM), in which the logarithm of the odds ratio is modelled via semi-parametric smoothing splines. A GAM has potential to uncover flexible relationships between the likelihood of disruption occurrence and confounding factors. The GAM with temporal correlation is presented in Equation (11):

$$e(X_{it}) = pr(W_{it} = 1 | X_{it} = x^{\{c\}}) = p(it), \tag{11}$$

$$log\left[\frac{p(it)}{1 - p(it)}\right] = \alpha + f(x^{\{c\}}; \beta) \quad i = 1, \dots, n \quad t = 1, \dots, T,$$

where $f(x^{\{c\}}; \beta)$ is a flexible spline function of baseline characteristics. After estimating propensity scores, we check the common support (overlap) assumption to ensure the effective matching and reliability of the propensity score estimates (Lechner, 2001).

The next step is matching. Every treated unit $i$ at time $t$ is paired with $M$ similar control units based on the value of their propensity scores and time-of-day characteristics. Since there is no theoretical consensus on the superiority of matching algorithms, we adopt two commonly used approaches: Subclassification Matching and Nearest Neighbour Matching. We then compare them with different replacement conditions and pairing ratios and select the one that balances the greatest disparity among the mean of confounding factors. It is also necessary to check the conditional independence assumption after matching. We conduct balancing tests to check whether the disrupted units and the matched units are statistically similar across the domain of confounders. If significant differences are found, we try another specification of the propensity score model and repeat the above-discussed procedure.

In the last step, we estimate station-level disruption impact using Equations (5) and (6). Given the matched pairs, the treatment effect for a station at a specific period is estimated as the difference between outcomes of the treated unit and its matched control units. Then the average station-level disruption impact is obtained by averaging these differences across all disrupted periods. We separately estimate the average treatment effects for three measures of metro performance:

1. *Entry ridership:* the number of passengers who enter the study unit.
2. *Average travel speed:* average of the speed of all trips that start from the study unit. For each trip, speed is computed as travel distance divided by observed journey time. Whereas journey time is directly obtained using the smart card data, travel distance (track length) of the most probable route is derived using the shortest path algorithm[5]. Passengers who had left the system and used other transport modes to reach the final destination are not included in the computation of this metrics. If the origin station is entirely closed and no passenger can continue trips by metro, then the average speed will be zero. If the origin station is partially closed, this metrics reflects the average speed of passengers who remain in the system.
3. *Distribution of passenger flow:* the distribution of completed trips that start from (outward flow) and arrive to (inward flow) the study units.

### 3.2 Stage 2: Constructing vulnerability metrics

We propose four station-level vulnerability metrics that are constructed from the empirical estimates of disruption impacts on the above-discussed performance measures.

i). The *loss of travel demand* is expressed as:

$$d_i = -\tau^i_{ATE}(entry), \qquad (12)$$

where $\tau^i_{ATE}(entry)$ (calculated using Equation 5) denotes the station-level change in the number of entry passengers due to service disruptions. $d_i$ is the loss of demand from external passengers who have not entered the metro system during a 15-minute interval due to disruption.

ii). The *loss of average travel speed* quantifies the decline in level of service experienced by each passenger at a metro station (individual delay), which is expressed as:

$$s_{avg}^i = \tau^i_{ATE}(speed), \qquad (13)$$

where $\tau^i_{ATE}(speed)$ (calculated using Equation 5) denotes the decrease in average travel speed of trips starting from station $i$ during a 15-minute disruption period. By definition, $s_{avg}^i$ accounts for the changes in both travel distance and journey time of passengers.

iii). The *loss of gross travel speed* reflects the loss of passenger kilometres per unit time, which is expressed as:

$$s_{gross}^i = \tau^i_{ATE}(speed) \times r_i, \qquad (14)$$

where $r_i$ denotes the average entry ridership of all disrupted 15-minute intervals at the corresponding station. Thus, $s_{gross}^i$ denotes the total decrease in average travel speed for all passengers who start their journeys from station $i$ during a 15-minute service disruption.

iv). The *irregularity in passenger flow* reflects the degree of deviation in the distribution of trips from/to the disrupted station as compared to regular conditions, which is expressed as:

---

[5] For future research, conditional on the availability of vehicle location data, the shortest path algorithm can be replaced by the passenger-train assignment algorithm (Hörcher, Graham and Anderson, 2015; Zhu and Goverde, 2019) to infer the most likely path chosen by passengers.

$$f_i = \tau^i_{ATE}(flow) \qquad (15)$$

where $\tau^i_{ATE}(flow)$ (calculated using Equation 6) denotes the average irregularity in flows that start from or arrive at station $i$ during a 15-minute disruption period. This metrics extends the scope of vulnerability measurement in terms of the entire distribution of entry/exit ridership, instead of just analysing the disruption impact on the entry or exit demand (that is, moments of the trip distribution).

**3.3 Stage 3: Imputing Missing Vulnerability Metrics**

Some stations may not encounter any incidents within the study period. Thus, the empirical disruption impact and the vulnerability metrics cannot be estimated directly for these stations. To predict the missing metrics of non-disrupted stations, we estimate a random forest regression model (Hastie et al. 2009):

$$\hat{f}^B_{rf}(x^{\{s\}}) = \frac{1}{B}\sum_{b=1}^{B} T(x^{\{s\}}; \theta_b), \qquad (16)$$

where $\hat{f}^B_{rf}(x^{\{s\}})$ denotes the random forest predictor. In the equation above, $B$ is the number of trees, $x^{\{s\}}$ is a vector of input features (see Table 2 for details). Furthermore, $T(x^{\{s\}}; \theta_b)$ is the output of the $b^{th}$ random forest tree, and $\theta_b$ characterizes the $b^{th}$ random forest tree. The random forest regression that we apply here is a combination of a bagging algorithm and ensemble learning techniques. By averaging the output of several trees (or weak learners in boosting terminology), it reduces the overfitting problem.

For this study, random forest (RF) is an appropriate prediction method. Interested readers are referred to Hastie et al. (2009) for details of RF regression algorithms, who explain the reasons behind its superior prediction accuracy as compared to other competing machine learning methods (Khalilia, Chakraborty and Popescu, 2011; Couronné, Probst and Boulesteix, 2018). However, considering that the field of machine learning is evolving rapidly, we also encourage readers to explore state-of-the-art alternatives to RF and test different prediction algorithms to find the most suitable algorithm for their data.

## 4. Case study: London Underground

In 2013, the London Underground (LU) had 270 stations and 11 lines, with a total length of 402 km stretching deep into Greater London. The circle-radial network structure, as shown in Figure 2 (Wikimedia Commons, 2013), is one of the largest and most complex metro systems in the world. Of all lines within the network, one is circular (Circle Line) covering Central London, and the remaining 10 are radial routes converging at the centre of the system. For connectivity among stations, LU has 56 stations connecting 2 lines, 16 stations connecting 3 lines and 8 stations connecting more than 4 lines. LU is also one of the busiest metro systems, with 1.265 billion journeys by the end of 2013 (Transport for London, 2019). Due to over 150 years old operations and enormous passenger demand, disruptions occur frequently in LU.

We use the following data to analyse the station-level vulnerability of the LU system. We conducted data processing and analysis using open-source R software (version 4.0.3).

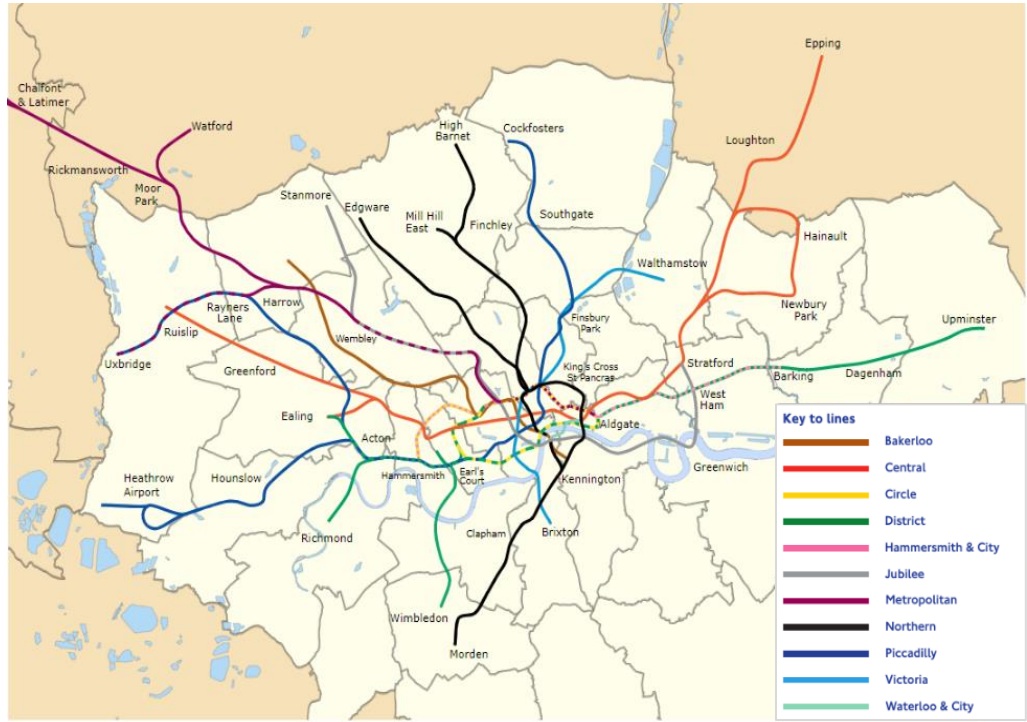

**Figure 2:** London Underground network [adapted from (Wikimedia Commons, 2013)].

*Pseudonymised smart card data:* Transport for London (TfL) provided automated fare collection data from 28/10/2013 to 13/12/2013 (35 weekdays) between 6:00 and 24:00. We consider this duration as our study period. The smart card data contain information on transaction date and time, entry and exit locations, encrypted card ID and ticket type (pay as you go/season ticket). The resolution of time stamps exacts to one minute. By using smart card data, we compute entry/exit ridership of each station and obtain passengers' journey time and travel speed.

*Incidents and service disruption information:* TfL also provided incident information data for our study period. By mining provided incidents logs, we construct an accurate database of service disruptions, which includes the occurrence time, location and duration of disruptions.

*LU network topology information:* We collect data on station coordinates, topology structure and the length of tracks between adjacent stations from open databases authorised by TfL[6].

*Weather data:* We collect temperature (°C), wind speed (km/h) and rain status from the Weather Underground web portal[7]. Based on the observations of over 1000 weather stations around London, we estimate weather conditions for all LU stations at 15-minute resolution for our study period.

*LU station characteristics:* These station-level features include daily ridership, station age, rolling stock age, sub-surface/deep-tube stations, terminal stations and screen doors. We also calculate supplementary factors, which capture the characteristics of the affected areas around metro stations. To compute these factors, we define the *affected area* as a circular area with the radius of 500 metres around the station. We use 2011 UK Census data at Lower Super Output Area (LSOA) level[8] to calculate these supplementary factors. We select all LSOAs whose centroids are within the 500 metres radius of the affected area. We then average the related statistics of the selected LSOAs according to their areas in the circle. Figure 3 illustrates the above process of calculations.

---

[6] Source: https://www.whatdotheyknow.com.
[7] Weather information web portal: https://www.wunderground.com/

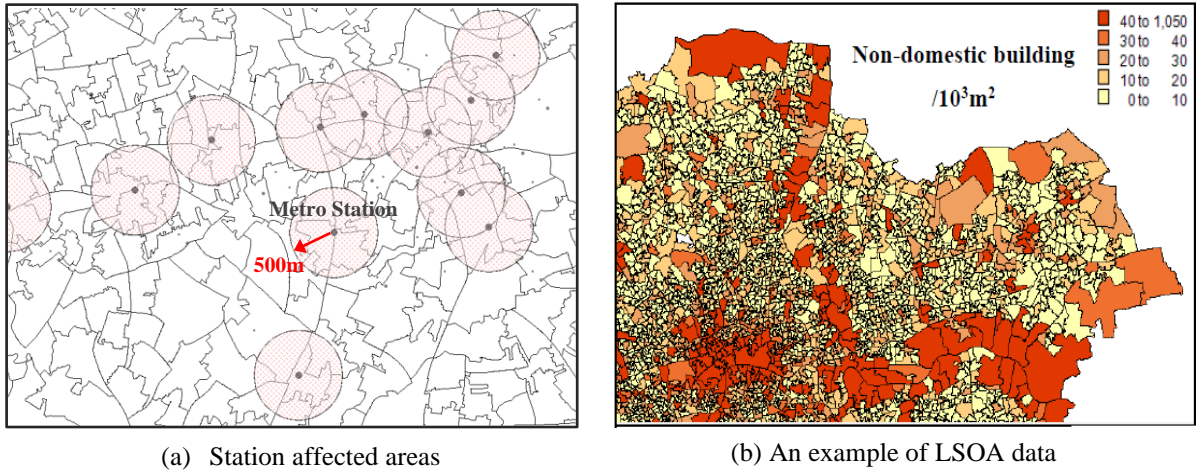

(a) Station affected areas    (b) An example of LSOA data

**Figure 3:** The illustration of calculating station-level supplementary factors.

To construct the causal inference framework for LU, our study unit is the observation of metro stations during each 15-minute interval within the system service time. We define *metro disruption* as the state when scheduled train services are interrupted for at least 10 minutes at a station. Over the study period, LU encountered 2894 disruptions lasting from 10 minutes to 11 hours. The aim of causal inference is to estimate the unbiased impact of these observed disruptions (i.e., treatment) on system-performance measures (outcome). The treatment status $W_{it}$ is constructed according to the disruption database mentioned in Section 4. To match the disruption duration with the timeframe of study units, we define the following rule to assign the treatment status: if a disruption occurs within a 15-minute interval $t$ of a given station $i$, we regard this study unit as disrupted (i.e., $W_{it} = 1$), no matter whether disruptions start or end in the middle or last for the entire 15-minute interval. Conversely, if the station is under normal service during entire 15-minute interval, we regard this study unit as un-disrupted (i.e., $W_{it} = 0$). The treatment outcomes $Y_{it}$ are presented as three station-level performance indicators: entry ridership, exit ridership, and average travel speed.

As discussed earlier, metro disruptions may not occur randomly. We list all potential confounding factors for LU in Table 2, which we use in estimating the propensity score model (Section 3.1). These confounders are selected according to the literature and expertise, including travel demand, weather conditions, engineering design, time of day and past disruptions (Brazil et al., 2017; Melo et al., 2011; Wan et al., 2015). Table 2 also shows available covariates for the imputation of missing vulnerability metrics in Stage 3 (Section 3.3), which not only include some of confounders, but also include supplementary factors of LU station characteristics.

---

[8] Source: London Datastore, published by Greater London Authority: https://data.london.gov.uk/census/.

Table 2: Available covariates for PSM and vulnerability imputation.

| Variable | Description | Stage 1 | Stage 3 |
|---|---|---|---|
| *Real-time travel demand* | | | |
| 15-minute entry ridership | The number of passengers that enter a station within 15 minutes before the study unit. | ✓ | |
| 15-minute exit ridership | The number of passengers that exit a station within 15 minutes before the study unit. | ✓ | |
| *Average travel demand and speed* | | | |
| Daily entry ridership | The daily average number of passengers that enter a station during the study period. | | ✓ |
| Daily exit ridership | The daily average number of passengers that exit a station during the study period. | | ✓ |
| Daily travel speed | The daily average speed of passengers that start their trips from the study unit. | | ✓ |
| *Weather conditions* | | | |
| Temperature | Atmospheric temperature around study units. Observations range from -3℃ to 20℃. | ✓ | |
| Wind speed | The wind speed around study units (km/h), ranges from 0 to 88 km/h. | ✓ | |
| Rain status | Dummy variable, representing whether it was raining at study units. | ✓ | |
| *Engineering design characteristics* | | | |
| Rail connectivity | Dummy variable, representing whether the station is connected to other rail systems. | ✓ | ✓ |
| Overground | Dummy variable, representing whether the station is on surface or closed deep in tube. | ✓ | ✓ |
| Terminal | Dummy variable, representing whether the station is an origin or terminal station. | ✓ | ✓ |
| Screen door | Dummy variable, representing whether the station has screen doors on the platform. | ✓ | ✓ |
| Number of lines | The number of lines within the given station, ranges from 1 to 6 in LU. | ✓ | ✓ |
| Average adjacent distance | The average distance between the given station and its adjacent stations (km). | ✓ | ✓ |
| Station age | Age of the oldest metro line served by the station. | ✓ | ✓ |
| Rolling stock age | Average age of all rolling stocks operated in the given station | ✓ | ✓ |
| Zone | Categorical variable, the zone where the station is located, ranges from 1 to 9 in London Underground. | ✓ | ✓ |

| | | |
|---|---|---|
| *Time of day* | Time of day divided into nine intervals; AM peak: 6:30 to 9:30, PM peak: 16:00 to 19:00 | ✓ |
| *Past disruptions* | | |
| Number of past disruptions occurred in the same day | Representation of the temporal correlation of disruption occurrence. | ✓ |
| **Station supplementary factors** | | |
| *Socio-economic characteristics* | | |
| Total population* | | ✓ |
| Number of employed people* | | ✓ |
| IMD* | Index of Multiple Deprivation scores | ✓ |
| *Land use characteristics* | | |
| Domestic buildings* | Area of domestic buildings ($10^3$ m$^2$) | ✓ |
| Non-domestic buildings* | Area of non-domestic buildings ($10^3$ m$^2$) | ✓ |
| Other land use* | Area of other land use ($10^3$ m$^2$) | ✓ |
| *Accessibility measures* | | |
| Number of bus stops* | | ✓ |
| Biking* | Sharing bicycle facility dummy | ✓ |
| Parking* | Car parking facility dummy | ✓ |
| Road area (m$^2$) * | | ✓ |
| Path area (m$^2$) * | | ✓ |

*computed for the affected area around each station

## 5. Results and Discussions

Out of 270 stations of the LU system, TfL provided the required datasets for 265 stations during the study period (28/10/2013 – 13/12/2013). Smart card data were missing for the remaining five stations. Our analysis only covers weekdays, during which the system is open for 18 hours per day, starting from 6:00 a.m. to midnight. Based on the assumption of exchangeability of weekdays (Silva et al., 2015), we generate a panel dataset with a total of 265×35×18×60/15=667,800 study units. Although the PSM method is a *data-hungry method*, the untreated pool (control group) is large enough to ensure adequate matches for treated units. Specifically, the ratio of the number of control and treatment units is around 15:1.

### 5.1 Propensity score models

We initially include three key baseline covariates – past disruptions, time of day and real-time travel demand – in the logistic regression. We then iteratively add one of the remaining covariates at a time from covariates listed in Table 2 and conduct the likelihood ratio test to decide whether the additional covariate should be included in the final specification or not. We also test Generalised Additive Models (GAM), but we do not observe any gains in the model fit. A high proportion of dummy variables (11 out of 19) may limit the gains from a flexible spline specification of the link function. The estimation results of the logistic regression model are summarised in Table 3.

The role of propensity score models is to establish a comprehensive index to represent all confounding factors, rather than predicting treatment assignment. While noting that the logistic regression model does not reveal the causal effect of covariates on the likelihood of incident occurrence, we succinctly discuss the multivariate correlations uncovered by this model. The coefficients of time dummies indicate that incidents are more likely to occur in morning peak hours. Positive signs on coefficients of the remaining confounders (except Rail dummy) confirm that all these factors increase the probability of encountering a disruption. Specifically, surface stations are more susceptible to the surrounding environment than those in tubes. We find statistically significant interaction effects between wind speed and Overground dummy. The accumulated number of past disruptions happened on the same day increases the probability of encountering another incident. Conclusively, the propensity score model reveals that the occurrence of metro disruptions is non-random, which, in turn, also justifies the application of causal inference methods in estimating disruption impacts.

Table 3: The results of propensity score model (logistic regression).

| Confounders | Coef. | S.E. |
| --- | --- | --- |
| Intercept | -4.547*** | 0.036 |
| Past disruptions | 0.271*** | 1.634e-03 |
| Time0 (6:00-6:30) (1) | 1.883*** | 0.027 |
| Time1 (6:30-7:45) (1) | 1.631*** | 0.021 |
| Time2 (7:45-8:45) (1) | 1.607*** | 0.022 |
| Time3 (8:45-9:30) (1) | 1.252*** | 0.026 |
| Time4 (9:30-16:00) (1) | 0.801*** | 0.016 |
| Time5 (16:00-17:15) (1) | 0.224*** | 0.026 |
| Time6 (17:15-18:15) (1) | 0.193*** | 0.028 |
| Time7 (18:15-19:00) (1) | 0.438*** | 0.029 |
| Temperature (°C) | 0.035*** | 1.926e-03 |
| Wind speed (km/h) | 0.017*** | 1.853e-03 |
| Rain (1) | 0.329*** | 0.015 |
| Rail (1) | -0.179*** | 0.013 |
| Overground (1) | 0.219*** | 0.023 |
| Ave distance (km) | 0.042*** | 4.748e-03 |
| Station age (max) | 5.714e-04** | 2.005e-04 |
| Pre 15-minute entry ridership | 1.969e-04*** | 2.098e-05 |
| Rolling stock age (mean) | 4.514e-03*** | 4.666e-04 |
| Overground*Wind speed | 0.014*** | 2.352e-03 |
| McFadden's pseudo R-squared | 0.184 | |

Note: (1) represents dummy variables
The base dummy for time of the day is Time8 (19:00-24:00).
*$p < 0.1$; **$p < 0.05$; ***$p < 0.01$.

Alternatively, the estimated propensity score model can also be viewed as a binary classifier that predicts whether metro disruptions occur or not. To illustrate its diagnostic ability, we compute the area under the receiver operating characteristic curve: AUC=0.796, which again indicates that the occurrence of metro disruptions is non-random.

## 5.2 Matching results

Before the estimated propensity scores are utilised for matching, we inspect the *common support* condition (assumption 2 of the PSM method). Figure 4 presents the propensity score distributions for both disrupted and normal observations. The histograms display apparent overlap between the treatment and control groups, even for large propensity scores. There is no treated unit outside the range of common support, which means we do not need to discard any observations. We thus conclude that the overlap assumption is tenable in our empirical study.

The PSM method aims to balance the distribution of confounders between the treatment and control groups after the matching stage. To assess the quality of matching, we perform balance tests for four algorithms: subclassification matching, nearest neighbour matching without replacement ($M = 1$), nearest neighbour matching with replacement ($M = 1$) and nearest neighbour matching with replacement ($M = 2$), where M is the number of matched control units for each treatment unit. It is worth noting that the proposed matching scheme not only conditions on the estimated propensity scores, but also condition on the time-of-day of the treatment (disruption). We find that nearest neighbour matching with replacement ($M = 2$) performs the best, improving the overall balance of all confounding factors by 99.95%. This improvement indicates that within matched pairs, the difference of propensity scores and time-of-day characteristics between treatment and control units has been reduced by 99.95%, compared with the original data before matching.

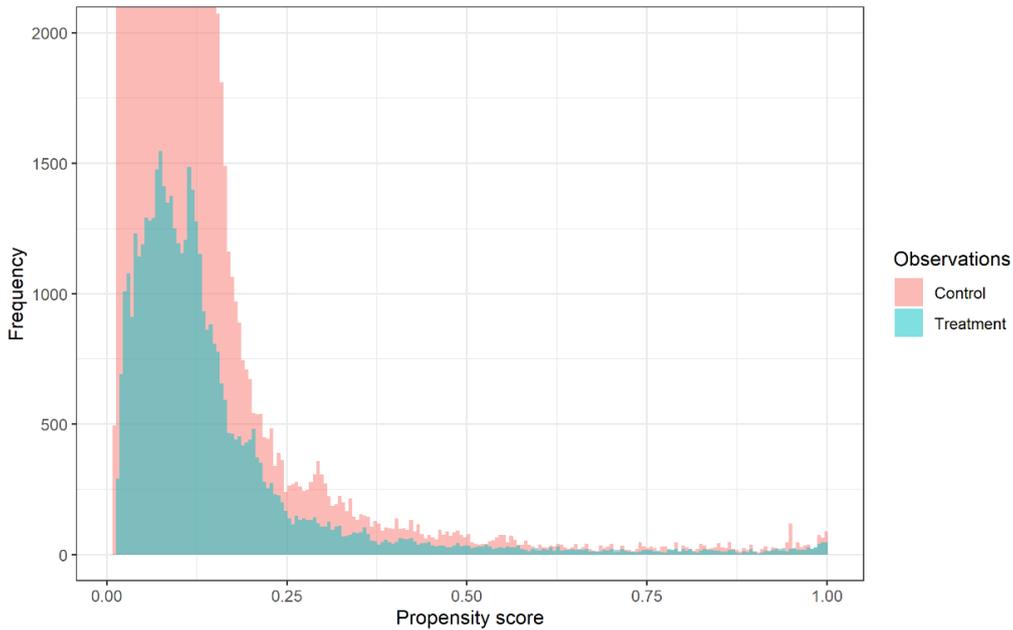

**Figure 4:** Histogram of propensity scores to test the Common Support condition[9].

---

[9] Due to higher share of the control group, the frequency in Figure 4 ranges up to 60,000 for lower propensity scores. However, we truncate frequency at 2,000 to clearly show the validity of overlap condition across the entire domain of the propensity score.

## 5.3 Imputation of missing vulnerability metrics

During the study period, 21 out of 265 stations did not encounter any service disruptions. We apply the random forest regression model to predict the missing vulnerability metrics of these stations. The input features of the model are indicated in "Stage 3" column of Table 2, consisting of station-level supplementary factors and a subset of confounding factors. For each vulnerability metrics, we estimate the random forest regression model using the 'randomForest' package of R (Liaw and Wiener, 2002). In terms of model settings, we consider the maximum number of trees to be 5000, randomly sample seven variables as candidates at each split, and assume the minimum size of terminal nodes to be two. The results show that more than 67% of the variance can be explained by input features for all vulnerability metrics. We summarize the prediction performance of random forest regression in Table 4 and benchmark it against two competing methods: linear regression and support vector machines.

**Table 4**: Prediction accuracy of different regression methods.

| Vulnerability metrics | Performance measures | Imputation methods | | |
|---|---|---|---|---|
| | | Random Forest | Linear Regression | Support Vector Machines |
| Demand loss | MAE | 2.794 | 33.089 | 5.181 |
| | RMSE | 4.285 | 37.342 | 9.766 |
| | RAE | 0.29 | 44.556 | 0.538 |
| | RSE | 0.095 | 330.181 | 0.493 |
| Avg. travel speed loss | MAE | 0.236 | 11.081 | 0.468 |
| | RMSE | 0.684 | 16.848 | 1.892 |
| | RAE | 0.318 | 1.151 | 0.63 |
| | RSE | 0.111 | 1.468 | 0.848 |
| Gross travel speed loss | MAE | 62.416 | 979.91 | 114.554 |
| | RMSE | 96.461 | 1224.723 | 216.472 |
| | RAE | 0.314 | 4.932 | 0.577 |
| | RSE | 0.107 | 17.18 | 0.537 |
| Irregularity in flow (Euclidean-entry) | MAE | 1.405 | 3.514 | 2.213 |
| | RMSE | 1.935 | 4.575 | 3.474 |
| | RAE | 0.23 | 0.574 | 0.362 |
| | RSE | 0.058 | 0.326 | 0.188 |
| Irregularity in flow (Hellinger-entry) | MAE | 0.02 | 0.051 | 0.034 |
| | RMSE | 0.025 | 0.064 | 0.048 |
| | RAE | 0.246 | 0.625 | 0.417 |
| | RSE | 0.066 | 0.418 | 0.234 |
| Irregularity in flow (KL-entry) | MAE | 0.276 | 0.498 | 0.333 |
| | RMSE | 0.184 | 0.72 | 0.613 |
| | RAE | 0.241 | 0.654 | 0.436 |
| | RSE | 0.074 | 0.506 | 0.366 |

Four measures are considered to benchmark the performance of random forest regression against other methods – mean absolute error (MAE), root mean squared error (RMSE), relative absolute error (RAE), and relative squared error (RSE). Whereas MAE measures the average magnitude of the errors in predictions, RMSE represents the standard deviation of the unexplained variance (Willmott and Matsuura, 2005). A better prediction model produces lower values of these performance measures. The results in Table 4 indicate that the random forest regression outperforms other competing methods with the lowest MAE, RMSE, RAE and RSE for all vulnerability metrics.

### 5.4 LU vulnerability metrics

The estimated vulnerability metrics vary across stations in the LU system. We first discuss results for loss of entry demand, loss of average travel speed, and loss of gross travel speed metrics. For 265 operated stations in 2013, during a 15-minute period of service disruption, the loss of station entry demand ranges from 0 to 103.4 passengers, the loss of average travel speed ranges from 0 to 21.76 kilometres/hour, and the loss of gross travel speed ranges from 0 to 2032.3 passenger-kilometres/hour. The spatial distributions of these vulnerability metrics are visualised in Figures 5(a) to 5(c). For the demand loss and gross speed loss, the large proportion of vulnerable stations are in inner London areas, while a small number of vulnerable stations are also located in suburban areas. Conversely, for the loss of average travel speed, the most vulnerable stations are scattered around outer London areas. These stations usually have only one metro line (internal alternatives) and have very limited access to other transport modes (external alternatives) compared to Central London areas. When passengers encounter disruptions, to continue their trips they need to wait for longer time in the system until train services are recovered. In other words, due to of lack of alternative routes[10], passengers at these stations tend to experience more individual delays.

We firstly sort all 265 stations based on demand and speed loss metrics, and the top 15 stations are presented in Table 5. Victoria is the most vulnerable station based on demand loss and gross speed loss metrics. Other stations such as Hammersmith, London Bridge, Kenton, Brixton are also among the top vulnerable stations based on both metrics. However, based on only the loss of average travel speed metrics, the most vulnerable stations are South Kenton, Kenton and North Wembley in outer London areas, where each passenger suffers the longest delay due to lack of alternative routes. The above rankings based on different vulnerability metrics can assist metro operators in preparing effective plans for ridership evacuation and service recovery.

Table 5 also presents normalised vulnerability metrics for these top 15 stations, which is the relative percentage change as compared to the undisrupted performance measure (baseline). Note that all baseline situations for these three metrics are calculated by using average across undisrupted observations. We find that the rankings based on relative vulnerability metrics can be different than those based on absolute metrics, especially for the loss of travel demand. In more isolated parts of the network, where alternative routes may not be available, stations can lose up to 136.4% of their normal demand due to service interruption (e.g. Kenton Station in Zone 4 with only no intersection metro line). This implies that more connected stations are actually less vulnerable in this respect, as passenger can find alternative routes if one of the lines becomes disrupted. This result also highlights potentially important distinctions in terms of the interpretation of the proposed metrics. In terms of relative metrics of average travel speed, the same top three vulnerable stations – South Kenton,

---

[10] There can be two types of alternative routes under disruptions – within the metro system (interchange to use other operated lines) and outside of it (in the form of other modes).

Kenton and North Wembley – experienced decrease in average travel speed by 108.9%, 92.8% and 36.1%, respectively, due to disruption. Kenton station is also the most vulnerable stations based on the relative loss of gross trip speed, which is reduced by 152.4%.

We propose three distance measures for the irregularity in flow metrics: Euclidean distance (ED), Hellinger distance (HD) and Kullback–Leibler (KL) divergence for both outward (from) and inward (to) flows. Euclidean distance directly compares the difference of each element of the trip distribution, where the *element* represents the ridership between a specific station and the disrupted station. ED reflects changes in the magnitude as well as the proportion of the flow of each element because it is not normalised. HD and KL divergence are normalised measures as they compare the difference between probability mass function of trip distributions, which capture only change in the proportion of trips completed between the disrupted and other stations. Unlike ED, HD and KL divergence would not be useful measures if disruption leads to a decrease in ridership across all stations by the same proportion. HD and KL divergence are close in principle, but the latter can be interpreted as the change in relative entropy, which is meaningful in the context of disruptions in metro systems. As an analogy with the concept of entropy in thermodynamics, we may interpret the extra entropy in metro systems as an additional *generalised cost* (in terms of time and congestion costs) that passengers have to pay under disruptions.

We plot the spatial distribution of all these distance measures in Figures 5(d) to 5(f). We also sort all 265 stations based on ED, HD and KL divergence, and the top 15 vulnerable stations are presented in Table 6. We find that the station rankings for outward flow (i.e., the entry ridership distribution) based on ED are similar to those obtained based on demand loss and gross speed loss metrics. They also share a similar spatial distribution of vulnerable stations. As for the distribution of inward flow (i.e., the exit ridership distribution), the most affected stations are mostly busy stations in Central London areas. As expected, station rankings based on HD and KL divergence are similar. For both inward (exit) and outward (entry) flow distributions, suburban stations are more severely affected than Central London stations on a normalised scale. The top 3 stations based on HD and KL divergence are South Kenton, Chesham and Heathrow Terminal 4.

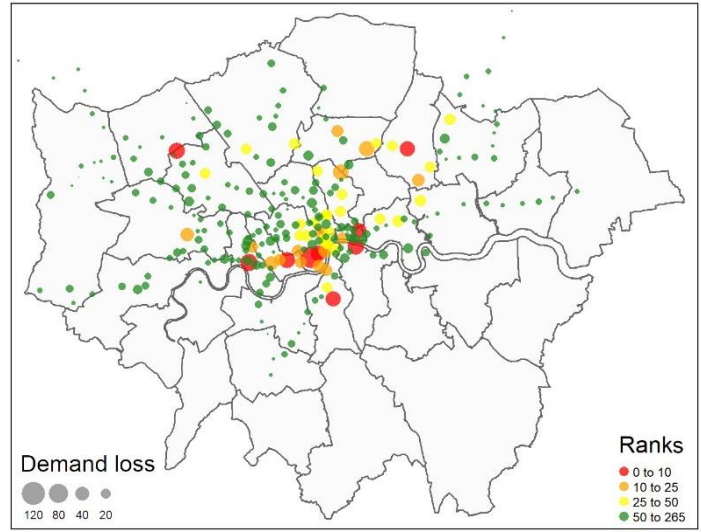

(a) The loss of travel demand

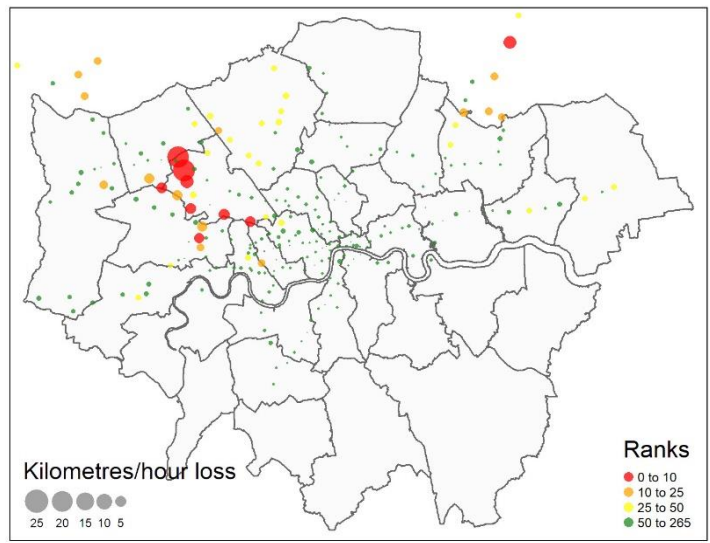

(b) The loss of average travel speed

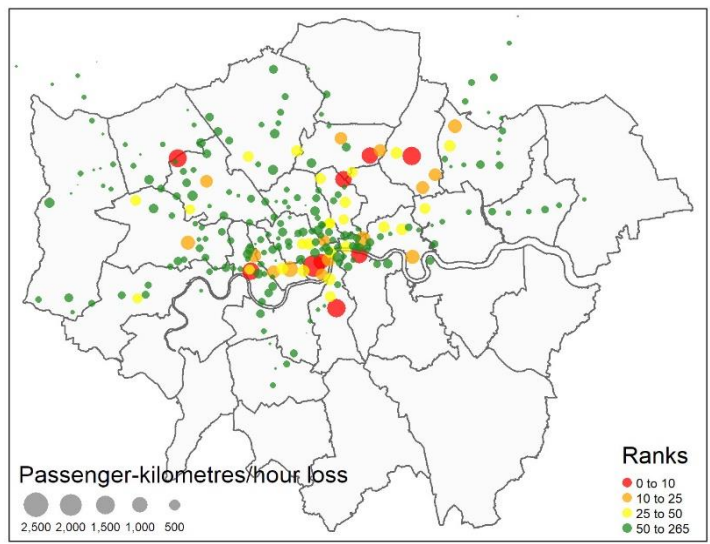

(c) The loss of gross travel speed

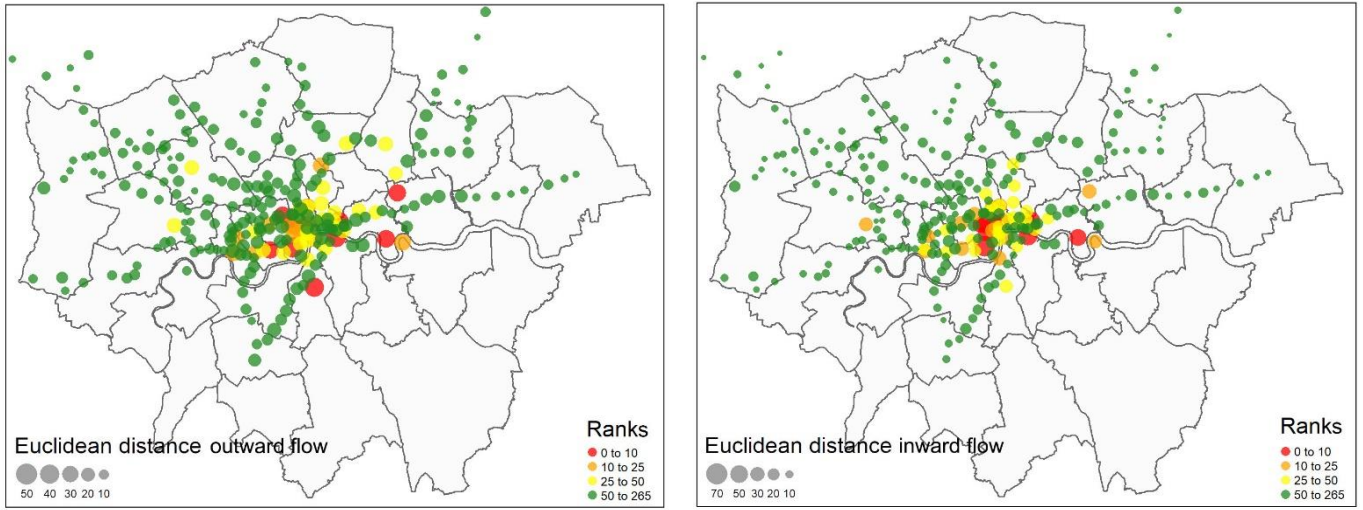

(d) The irregularity in flow distribution from/to the disrupted station (Euclidean distance)

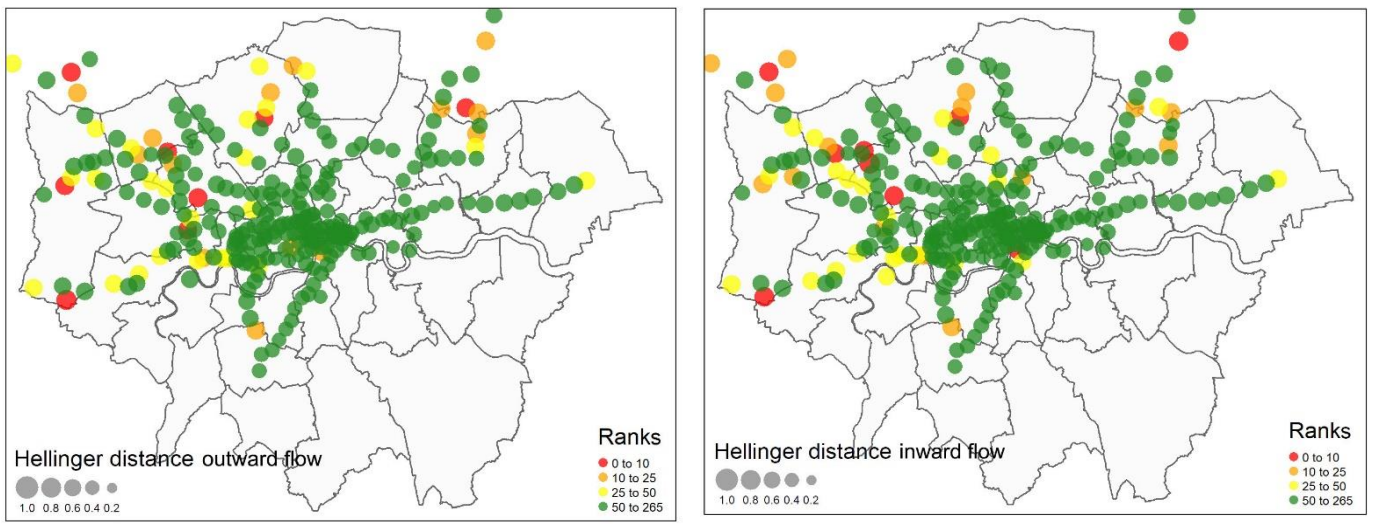

(e) The irregularity in flow distribution from/to the disrupted station (Hellinger distance)

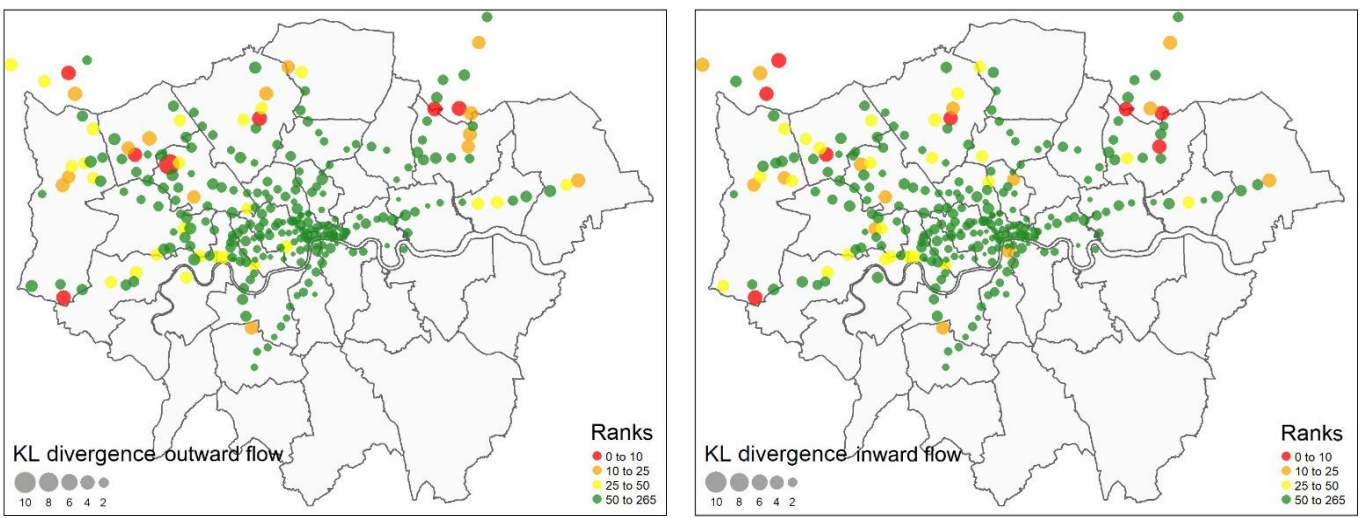

(f) The irregularity in flow distribution from/to the disrupted station (KL divergence)

**Figure 5:** Spatial distribution of station-level vulnerability metrics in London Underground.

**Table 5:** Top 15 vulnerable stations based on demand loss and speed loss vulnerability metrics.

| Station | Demand loss in passenger/15-minute (% of baseline) | Station | Avg. travel speed loss in km/h (% of baseline) | Station | Gross travel speed loss in passenger-km/h (% of baseline) |
|---|---|---|---|---|---|
| Victoria | 103.4 (13.0%) | South Kenton | 21.76 (108.9%) | Victoria | 2032.3 (13.6%) |
| Hammersmith | 66.8 (15.4%) | Kenton | 20.35 (92.8%) | Walthamstow Central | 1480.2 (18.6%) |
| London Bridge | 60.0 (8.6%) | North Wembley | 7.10 (36.1%) | Brixton | 1439.0 (12.4%) |
| South Kensington | 59.8 (12.3%) | Theydon Bois | 6.72 (23.3%) | Kenton | 1376.9 (152.4%) |
| Kenton | 58.8 (136.4%) | Harlesden | 5.22 (24.4%) | Hammersmith | 1326.9 (15.0%) |
| St. James's Park | 56.6 (22.3%) | Kensal Green | 4.64 (24.0%) | Seven Sisters | 1176.0 (16.6%) |
| Brixton | 53.8 (11.0%) | Alperton | 4.62 (21.4%) | London Bridge | 1152.6 (8.7%) |
| Liverpool Street | 52.2 (6.7%) | Sudbury Hill | 4.54 (21.2%) | Finsbury Park | 1148.9 (14.0%) |
| Walthamstow Central | 51.4 (17.8%) | North Ealing | 4.45 (21.9%) | St. James's Park | 1015.9 (21.9%) |
| Finsbury Park | 48.5 (14.2%) | South Harrow | 4.44 (20.2%) | Liverpool Street | 993.3 (6.8%) |
| Seven Sisters | 47.9 (18.3%) | Sudbury Town | 4.26 (19.0%) | South Kensington | 978.1 (11.3%) |
| Earl's Court | 41.3 (12.9%) | Park Royal | 4.20 (20.4%) | Ealing Broadway | 813.5 (12.7%) |
| Westminster | 39.3 (13.4%) | Roding Valley | 3.27 (20.6%) | Canary Wharf | 771.7 (4.7%) |
| Tottenham Court Road | 37.8 (6.8%) | Ruislip Gardens | 3.08 (12.5%) | Westminster | 755.7 (14.2%) |
| Ealing Broadway | 36.6 (12.6%) | Moor Park | 2.90 (11.0%) | Shepherd's Bush | 727.9 (10.1%) |

**Table 6:** Top 15 vulnerable stations based on irregularity in flow vulnerability metrics.

| Station | ED (outward) | Station | ED (inward) | Station | HD (outward) | Station | HD (inward) | Station | KL (outward) | Station | KL (inward) |
|---|---|---|---|---|---|---|---|---|---|---|---|
| Victoria | 46.92 | Victoria | 66.95 | Chesham | 0.89 | South Kenton | 0.89 | South Kenton | 9.20 | Chesham | 9.77 |
| Liverpool Street | 44.36 | Oxford Circus | 66.84 | Heathrow Terminal 4 | 0.81 | Heathrow Terminal 4 | 0.82 | Chesham | 7.70 | Heathrow Terminal 4 | 5.00 |
| London Bridge | 42.82 | London Bridge | 56.85 | West Finchley | 0.76 | West Harrow | 0.78 | Heathrow Terminal 4 | 4.64 | Grange Hill | 4.95 |
| Oxford Circus | 41.79 | Liverpool Street | 54.61 | Croxley | 0.75 | West Finchley | 0.77 | Croxley | 4.64 | West Harrow | 4.84 |
| Brixton | 41.57 | Canary Wharf | 44.86 | Stonebridge Park | 0.74 | Croxley | 0.76 | Roding Valley | 4.53 | Watford | 4.56 |
| Canary Wharf | 35.77 | Holborn | 39.49 | Kenton | 0.73 | Stonebridge Park | 0.76 | West Finchley | 4.47 | West Finchley | 4.45 |
| Stratford | 33.76 | Tottenham Court Road | 39.16 | North Ealing | 0.73 | Theydon Bois | 0.76 | Amersham | 4.29 | Barkingside | 4.43 |
| South Kensington | 33.58 | Green Park | 35.94 | Chigwell | 0.73 | Kenton | 0.76 | Chigwell | 4.29 | Roding Valley | 4.39 |
| Baker Street | 32.87 | Bond Street | 34.98 | Hillingdon | 0.72 | Lambeth North | 0.75 | West Harrow | 4.18 | Moor Park | 4.36 |
| Hammersmith | 31.98 | Hammersmith | 34.90 | Amersham | 0.72 | North Ealing | 0.74 | Grange Hill | 4.14 | Croxley | 4.23 |
| King's Cross | 30.11 | Waterloo | 34.63 | West Harrow | 0.72 | Moor Park | 0.74 | Moor Park | 4.12 | Theydon Bois | 4.15 |
| North Greenwich | 30.03 | South Kensington | 34.39 | Moor Park | 0.72 | Barkingside | 0.74 | Harrow & Wealdstone | 4.00 | Lambeth North | 4.08 |
| Shepherd's Bush | 30.00 | Euston | 34.36 | Lambeth North | 0.72 | Grange Hill | 0.74 | Chalfont & Latimer | 3.96 | South Kenton | 4.06 |
| Finsbury Park | 29.78 | Leicester Square | 34.26 | Hyde Park Corner | 0.71 | Woodside Park | 0.73 | Upminster Bridge | 3.93 | Chigwell | 4.05 |
| Leicester Square | 29.39 | Vauxhall | 33.68 | Chesham | 0.71 | Wimbledon Park | 0.73 | Hillingdon | 3.87 | Chorleywood | 3.98 |

**Note:** ED: Euclidean distance, HD: Hellinger distance, KL: Kullback–Leibler divergence.

# 6. Conclusions and Future Work

Incidents occur frequently in urban metro systems, causing delays, crowding and substantial loss of social welfare. Operators need accurate estimates of vulnerability measures to identify the bottlenecks in the network. We propose a novel causal inference framework to estimate station-level vulnerability metrics in urban mero systems and empirically validate it for the London Underground system. In contrast to previous simulation-based studies, which largely assume virtual incident scenarios and necessitate the adoption of unrealistic assumptions on passenger behaviour, our approach relies on real incident data and avoids making behavioural assumptions by leveraging automated fare collection (smart card) data. We also illustrate that incidents can occur non-randomly, which further justifies the importance of the proposed causal inference framework in obtaining the unbiased estimate of disruption impacts.

The proposed empirical framework consists of three stages. First, we conduct propensity score matching methods and estimate unbiased disruption impacts at the station level. The estimated impacts are subsequently used to establish vulnerability metrics. In the last stage, for non-disrupted stations, we impute their vulnerability metrics by using the random forest regression model. We propose three empirical vulnerability metrics at station level, which are loss of travel demand, loss of average travel speed and loss of gross travel speed. The demand loss metrics reflects the amount of passenger who i) switched to other transport modes, ii) switched their departure time, trip origin or destination, iii) ended their trip, before entering the disrupted metro system. In other words, it implies the demand for alternative transport services during disruptions, which can guide metro operators to prepare effective service replacement plans. The two speed related metrics reflect the degradation in the level of service for passengers who still use the metro system under disruptions. These metrics provide essential information for service recovery to mitigate the adverse influence on passengers and the overall performance of stations. The proposed irregularity in flow metrics extends the scope of vulnerability measurement to the changes in trip distribution. This irregularity metrics can be used to reflect the level of disorder within metro systems.

The results of the case study of London Underground in 2013 indicate that the effect of service disruption is heterogeneous across metro stations and it depends on the location of a station in the network and other station-level characteristics. In terms of the travel demand loss and gross speed loss (overall delay), the most affected stations are more likely to be found in Central London areas, such as Victoria, London Bridge and Liverpool Street. On the other hand, considering average speed loss (individual delay), the most affected stations are scattered around outer London areas (e.g., South Kenton and Kenton) due to lack of alternative routes.

Disruption impact estimates are probabilistic relative to the sample data, that is, causal estimates and vulnerability metrics estimates have sampling distribution. Since our analysis is based on the data of LU from October 28 to December 13, 2013, the results of our case study reflect the vulnerability status of LU for this specific period. If we use data from other periods, the estimates of vulnerability metrics might change due to inherent temporal variations in travel demand and incidents. Therefore, to improve the generalisability of vulnerability metrics estimates, the study period needs to be long enough such that the sample is representative of the population. That is, a sample should capture supply-side interruptions as much as possible, including service disruptions due to maintenance. In addition, the sample should also reflect the possible fluctuations of travel demand.

The proposed methodology to obtain the unbiased estimates of disruption impact thus provides crucial information to metro operators for disruption management. It helps in identifying the bottlenecks in the network and in preparing targeted plans to evacuate ridership as well as to recover services in case of incidents. The direct integration of the estimated vulnerability metrics in preparing these target plans remains an avenue for future research. It is worth noting that the proposed framework can be applied to other metro systems conditional on the availability of the required data on incident logs, confounding characteristics and performance outcomes. Future empirical studies can also incorporate other context-specific and relevant confounders or outcome indicators in their analysis. For example, they can explore the disruption impacts on interchange passengers if the required datasets are available. We do not include this part of ridership in our LU case study because it cannot be directly derived from smart card data. More advanced assignment algorithm is required to identify passengers' routes by matching smart card data with vehicle location data and reproduce the spatiotemporal flow distribution in the metro network.

In line with the limitations of this study, there are three potential directions for future research. First, stations surrounding the disrupted stations may also be affected due to indirect propagation, but this study does not account for such spillover effects. Modelling spatiotemporal propagation disruption impacts requires significant methodological developments, which would be an important improvement over the current method. For instance, recent developments in Bayesian nonparametric sparse vector autoregressive models (Billio, Casarin and Rossini, 2019) can be adapted to model the spatiotemporal effect of service disruptions in transit networks. Second, the proposed vulnerability metrics can reveal static disruption impacts at different stations, but passengers need real-time service information to reschedule their trips. Thus, the current framework can be extended to update the vulnerability metrics dynamically. Considering the interaction between information provision and how it influences passengers' decision under disruptions, this advancement would improve the dissemination of the incident alerts to passengers in real-time. Finally, by merging data from other travel modes (e.g., bus, urban rails, shared bike or taxi) with metro datasets, we can estimate multi-modal vulnerability metrics in the same causal inference framework and understand the characteristics of the mode shift due to disruptions. In a potential extension of our method to multi-modal transport systems, the lost demand would not include passengers who shift to other public transport modes due to metro disruptions. Compared to metro-only vulnerability metrics, multi-modal demand loss metrics would focus on passengers who give up their trips entirely or switch to private transport modes. Therefore, for metro stations linked to multi-modal hubs, the multi-model demand loss metrics might be lower than the metro-only metrics. The magnitude of this gap would depend on the attractiveness of alternative public transport services compared to private modes.